\documentclass{article}
\usepackage{spconf,amsmath,graphicx,hyperref}
\usepackage{amssymb}
\usepackage{multirow} 
\usepackage{array} 

\title{Toward Human-Aligned Judgement of Speech Emotion Similarity}
\name{
\begin{tabular}{c}
    Yun-Shao Tsai$^1$, 
    Yi-Cheng Lin$^{1,*}$, 
    Chih-Kai Yang$^{1,*}$, 
    Ho-Jung Cheng$^2$, 
    Tsun-Yi Chang$^3$, \\
    Sheng-Wei Wu$^4$, 
    Yi-Shan Chen$^4$, 
    Hsiang-Chun Chang$^3$, 
    Liang-Chieh Lee$^3$, 
    Hung-yi Lee$^{1,3,5}$%
\end{tabular}%
  \thanks{$^*$Equal second-author contribution.}
}
\address{
  $^1$Graduate Institute of Communication Engineering, National Taiwan University, Taiwan \\
  $^2$Independent Researcher \qquad $^3$National Taiwan University, Taiwan \\
  $^4$National Taiwan University of Science and Technology, Taiwan \\
  $^5$NTU Artificial Intelligence Center of Research Excellence (NTU AI-CoRE), Taiwan
}
\begin{document}
\ninept
\maketitle
\begin{abstract}
Evaluating emotion preservation in expressive speech generation involves assessing how closely generated speech matches a reference in emotion.
Human listening tests assess this similarity, but their cost motivates automatic measures aligned with human judgments.
To support the development and evaluation of such measures, we introduce SES-Bench, a speech emotion similarity benchmark built from human comparisons of two candidate utterances against a shared reference.
These comparisons record which candidate listeners find emotionally closer to the reference and the strength of their preference.
Using these annotations, we train SES-Judge to score emotion similarity between two utterances.
SES-Judge significantly outperforms embedding cosine similarity and prompted large audio-language models in preference accuracy and correlation with human ratings that capture both preference direction and strength.
\end{abstract}
\section{Introduction}
\label{sec:intro}
Recent advances in speech generation have enabled increasingly expressive outputs in text-to-speech (TTS), voice conversion (VC), and speech-to-speech translation (ST)~\cite{zhou2026indextts2,su25c_interspeech, chen2026move}.
Evaluating these systems requires assessing how well outputs convey intended emotion.
For systems using a reference utterance to guide emotional expression, this means assessing how closely each output matches the reference in emotion.
Human listening tests assess this similarity, but their cost makes repeated evaluation during model development difficult~\cite{tsai2026false}.
This motivates automatic measures to assess emotion similarity at scale while remaining aligned with human perception~\cite{saeki22c_interspeech, ren2025highratemos}.

A common approach is EMO SIM, which computes cosine similarity between reference and generated speech embeddings, commonly extracted using emotion2vec~\cite{ma-etal-2024-emotion2vec,11463429}.
However, these scores do not consistently agree with human choices~\cite{tsai2026false}.
To learn and evaluate measures aligned with human perception, we need comparisons that capture both which candidate sounds emotionally closer to a reference and how strongly listeners prefer it.

To support this approach, we introduce SES-Bench, a Speech Emotion Similarity Benchmark for evaluating how well automatic similarity measures agree with human judgments.
Each sample is a triplet comprising one reference utterance and two generated candidate utterances.
Five annotators independently compare the candidates against the reference, using a seven-level scale to indicate which candidate sounds emotionally closer and how strongly they prefer it.
These annotations support learning a similarity measure and evaluating agreement with human perception.
We use these annotations to train SES-Judge as a Speech Emotion Similarity Judge that assesses how similar two utterances sound in emotional expression.
Our evaluation measures agreement with human choices and correlation between its score differences and human ratings.
SES-Judge significantly improves both measures over embedding similarity and judgments from large audio-language models (LALMs).
\begin{itemize}
\item We introduce SES-Bench\footnote{\url{https://github.com/danielqwer/Speech-Emotion-Similarity.git}}, a dataset and benchmark for human-perceived speech emotion similarity.
It provides graded human comparisons for model training and a shared test set for evaluating agreement with human judgments.
\item We develop SES-Judge, which learns from these graded comparisons to assess emotional similarity between two utterances.
It significantly outperforms pretrained embedding similarity and LALM judges in both preference accuracy and correlation with human ratings.
\end{itemize}

\section{Related Work}
\label{sec:related}

\subsection{Speech Emotion Datasets}
Speech emotion and expressive speech datasets describe utterances through categorical or dimensional labels~\cite{11457331,6849440}, or natural-language descriptions of speaking style that extend beyond fixed labels~\cite{jin2024speechcraft}.
However, similarity rankings derived from categorical or dimensional emotion labels can disagree with annotators' choices~\cite{9612052}.
For generated speech, prior work asks listeners to judge \emph{which} candidate is emotionally closer to a reference, but does not record \emph{how much} closer they perceive it to be~\cite{tsai2026false}.
To capture this degree of perceived difference, SES-Bench uses a graded comparison scale that distinguishes a slight similarity advantage from a large one, providing annotations for both training and evaluation.

\subsection{Emotion Similarity Evaluation Methods}
Emotion similarity is commonly measured using cosine similarity between utterance embeddings~\cite{11463429}.
Beyond using pretrained embeddings, similarity functions can be learned from triplets constructed using emotion labels~\cite{9612052}.
Human ratings also provide supervision for AutoPCP, which predicts overall prosodic similarity between speech pairs~\cite{barrault2023seamless}.
Speech judge models also learn from human feedback, although they assess naturalness and overall quality rather than emotional similarity to a reference~\cite{ICLR2026_8acd02d0,wang-etal-2026-speechllm}.
In concurrent work, STEB evaluates emotion preservation in speech translation using an LLM to compare emotion descriptions generated from the source and translated speech~\cite{cheng2026steb}.
SES-Judge learns a reference--candidate emotion similarity score from SES-Bench's graded human comparisons.
These comparisons supervise the score differences, encouraging them to reflect both annotators' preferred candidate and the perceived degree of difference.

\section{SES-Bench}
\label{sec:dataset}

\subsection{Task Definition}
We define perceived emotion similarity as the degree to which two utterances resemble each other in emotional expression.
In speech generation evaluation, we treat the utterance providing the desired expression as a reference $r$ and a generated utterance as a candidate $x$.
A judge assigns a score $s(r,x)\in\mathbb{R}$, where higher values indicate closer emotion matches.

SES-Bench collects relative similarity judgments using triplets comprising one reference $r$ and two candidates $x_A$ and $x_B$.
Each reference--candidate pair is scored independently, and the difference between the two scores represents the candidates' relative similarity to the reference:
\begin{equation}
\Delta(r,x_A,x_B)=s(r,x_A)-s(r,x_B).
\label{eq:score_difference}
\end{equation}
Positive values of $\Delta$ favor A, while negative values favor B.
The magnitude $|\Delta|$ represents the predicted strength of this difference.

\subsection{Data Construction}
Candidate utterances are generated using zero-shot TTS and VC, with each triplet pairing outputs from either two TTS systems or two VC systems.
To evaluate expressive speech generation, we exclude references labeled solely as neutral from both partitions.
For TTS, the two systems receive the same reference utterance and target text.
We generate this text using Claude Opus 5.\footnote{\url{https://www.anthropic.com/news/claude-opus-5}}
For VC, the source utterance serves as the reference, and the systems convert its speaker identity while retaining its linguistic content.
Both systems use the same target speaker prompt to specify the speaker identity.

The training set uses MSP-Podcast utterances~\cite{11457331} as references.
We generate TTS candidates using CosyVoice 3~\cite{du2025cosyvoice}, Step-Audio-EditX~\cite{yan2025step}, F5-TTS~\cite{chen-etal-2025-f5}, and IndexTTS2~\cite{zhou2026indextts2}, and VC candidates using AdaptVC~\cite{10889396}, EZ-VC~\cite{joglekar-etal-2025-ez}, Vevo2~\cite{11425816}, and FreeVC~\cite{10095191}.
The TTS systems include autoregressive and non-autoregressive models, while the VC systems include VITS-based and flow-matching architectures.
We randomly sample candidate pairs from these outputs, keeping each system approximately equally represented.

The test set uses CREMA-D utterances~\cite{6849440} as references to evaluate on a different corpus.
To test generalization to unseen generation systems, we use Qwen3-TTS~\cite{hu2026qwen3} for TTS and Seed-VC~\cite{liu2024zero} for VC exclusively in the test set.
Each test triplet pairs an output from one of these systems with an output from a randomly selected system used to generate training candidates for the same task.

\begin{figure}[t]
\centering
\includegraphics[width=0.92\columnwidth]{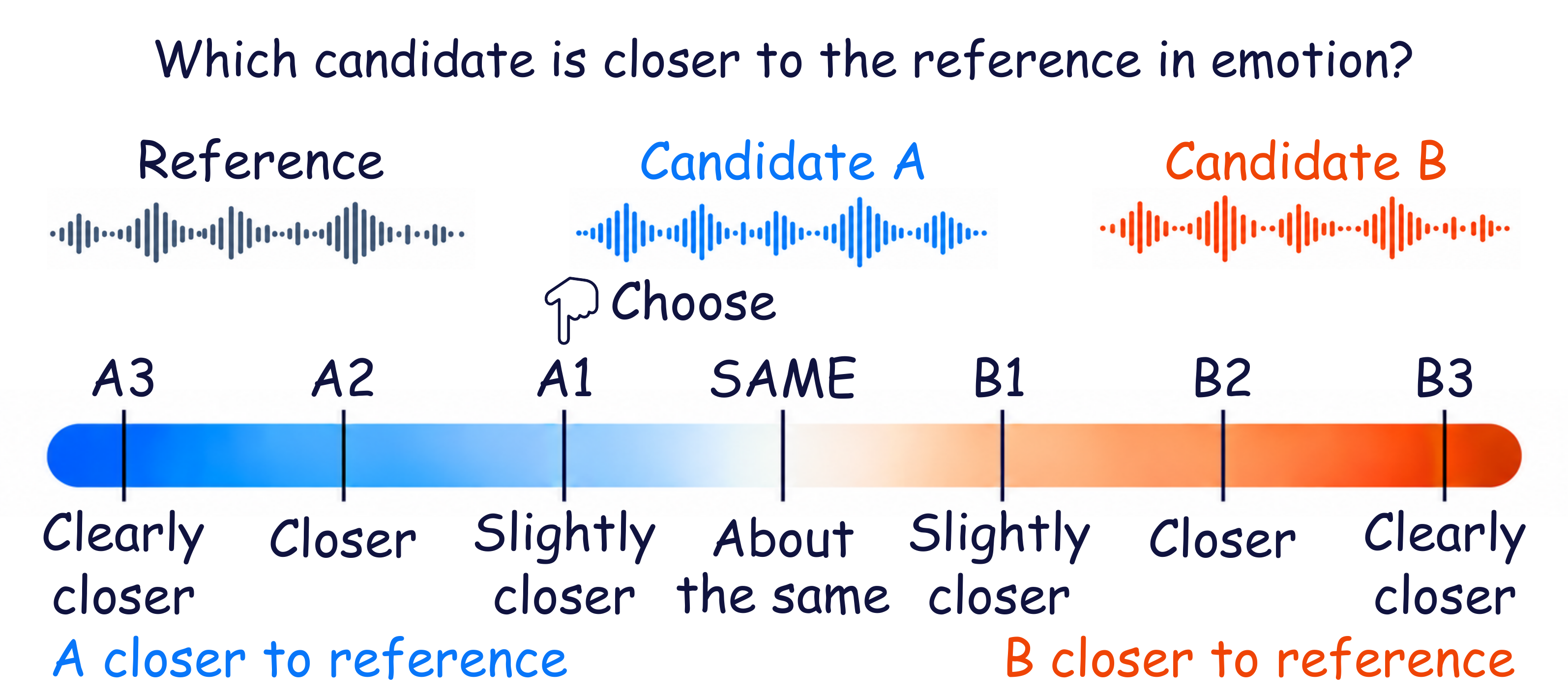}
\caption{Annotation schematic.
An annotator compares two candidates with a reference and rates their relative emotional similarity on a seven-point scale.}
\vspace{-3mm}
\label{fig:annotation_protocol}
\end{figure}

\subsection{Annotation and Dataset Statistics}
\label{sec:annotation}
As shown in Fig.~\ref{fig:annotation_protocol}, five annotators independently listen to each triplet and select one of the seven ordered responses A3, A2, A1, SAME, B1, B2, and B3.
A1, A2, and A3 mean that A is ``slightly closer,'' ``closer,'' and ``clearly closer'' to the reference in emotion, respectively, with corresponding meanings for B.
The numbered levels indicate the strength of this preference rather than emotional intensity.
SAME indicates approximately equal similarity to the reference, even if neither candidate matches it closely.

\begin{figure*}[t]
\centering
\includegraphics[width=\textwidth]{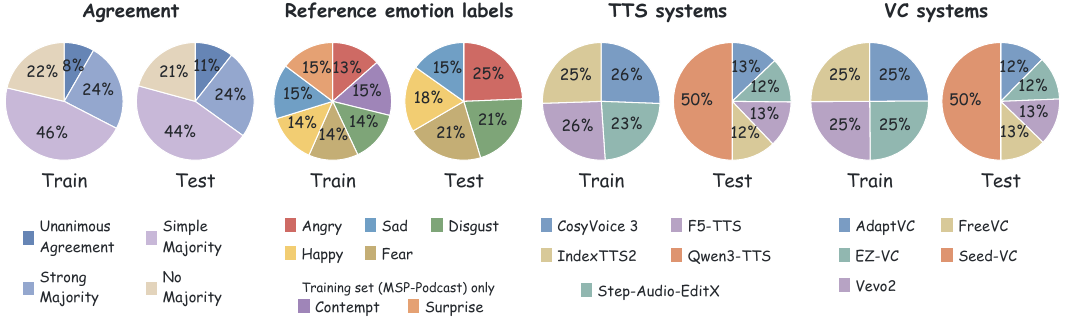}
\vspace{-3mm}
\caption{Distributions of agreement, reference emotion labels, and generation systems in SES-Bench, with Train and Test shown side by side for each group.
System proportions count candidate occurrences separately for TTS and VC.}
\vspace{-3mm}
\label{fig:dataset_distributions}
\end{figure*}

We remove any triplet flagged by at least one annotator as containing corrupted or unplayable audio.
This leaves 2,974 training comparisons (1,491 TTS and 1,483 VC) and 848 test comparisons (417 TTS and 431 VC).
The reference utterances cover 973 speakers in training (54.0\% male, 46.0\% female) and 91 in testing (52.7\% male, 47.3\% female).
The 11,466 reference and candidate utterances total 10.97 hours, with 8.86 hours in training and 2.11 hours in testing.
We summarize agreement among annotators by grouping A1--A3 as A and B1--B3 as B, with SAME as a third category.
The proportion of annotators supporting the most frequent category defines Unanimous Agreement ($5/5$), Strong Majority ($4/5$), Simple Majority ($3/5$), or No Majority ($2/5$).
Figure~\ref{fig:dataset_distributions} summarizes distributions of agreement levels, reference emotions, and generation systems for the training and test sets.
After removing comparisons flagged for corrupted or unplayable audio, more test comparisons have A as the majority answer than B.
To reduce this imbalance, we randomly select some comparisons with answer A and exchange candidates A and B, updating the annotations accordingly.
Among comparisons with Simple Majority or higher agreement, the resulting answer distribution is 318 A, 317 B, and 37 SAME.
This changes only candidate positions, preserving the audio pairs and human judgments.

\begin{table*}[t]
\centering
\fontsize{9}{10.5}\selectfont
\setlength{\tabcolsep}{2.5pt}
\renewcommand{\arraystretch}{1.12}
\caption{Baseline configurations.
Training data and objectives refer to additional training in this work.
Each listed scoring input defines a separate evaluation.
Fused embeddings combine layer representations; head-hidden embeddings are taken after the downstream hidden layer.
Classification is trained separately with and without synthetic speech.}
\label{tab:baseline_summary}
\begin{tabular}{@{}>{\raggedright\arraybackslash}m{0.16\textwidth}>{\raggedright\arraybackslash}m{0.13\textwidth}>{\raggedright\arraybackslash}m{0.18\textwidth}>{\raggedright\arraybackslash}m{0.25\textwidth}>{\raggedright\arraybackslash}m{0.23\textwidth}@{}}
\hline
Model / Encoder & Training data & Training task / loss & Scoring input & Scoring rule \\
\hline
EmotionRankCLAP & --- & --- & Projected audio embedding & Cosine similarity \\
WavLM-SER & --- & --- & SER head-hidden embedding & Cosine similarity \\
emotion2vec & --- & --- & Final encoder-layer embedding & Cosine similarity \\
\hline
\multirow{3}{=}{WavLM-SER\par emotion2vec}
& \multirow{3}{=}{MSP-Podcast\par with / without\par synthetic speech}
& \multirow{3}{=}{Emotion classification\par Cross-entropy}
& Fused embedding & Cosine similarity \\
& & & Head-hidden embedding & Cosine similarity \\
& & & Class probabilities & Negative KL divergence \\
\hline
\multirow{3}{=}{WavLM-SER\par emotion2vec}
& \multirow{3}{=}{MSP-Podcast}
& \multirow{3}{=}{VAD regression\par MSE}
& Fused embedding & Cosine similarity \\
& & & Head-hidden embedding & Cosine similarity \\
& & & Predicted VAD values & Negative Euclidean distance \\
\hline
WavLM-SER\par emotion2vec & MSP-Podcast & VAD-based similarity\par Triplet loss & Learned embedding & Cosine similarity \\
\hline
Qwen3-Omni\par Gemini 3.8 Flash & --- & --- & Seven-option responses & Mean signed rating / Most frequent A/B/SAME category \\
\hline
\end{tabular}
\end{table*}

\section{SES-Judge}
\label{sec:model}
We use a pretrained, frozen WavLM-based dimensional speech emotion recognition model (WavLM-SER)~\cite{goncalves24_odyssey} as our audio encoder.\footnote{\url{https://huggingface.co/3loi/SER-Odyssey-Baseline-WavLM-Multi-Attributes}}
We temporally average frame-level features to obtain utterance representations, apply a separate trainable linear projection~\cite{wang24y_interspeech} and normalization to each layer representation, and combine them through a learnable weighted sum.
A final linear projection produces a 1024-dimensional embedding, and we use cosine similarity to score each reference--candidate pair.

We train the model on SES-Bench's graded comparisons using cumulative logit ordinal regression.
The difference between the two similarity scores is mapped to probabilities over seven responses ordered from B3 through SAME to A3 using a learned positive scale and ordered thresholds.
We constrain the thresholds to be symmetric about zero so that exchanging A and B reverses the predicted response distribution.
We minimize cross-entropy against the empirical distribution of all five annotators' responses:
\begin{equation}
\mathcal{L}_{\mathrm{ord}}=-\frac{1}{N}\sum_{i=1}^{N}\sum_{k=1}^{7}q_{ik}\log p_{ik},
\label{eq:ordinal_loss}
\end{equation}
where $N$ is the number of triplets, $q_{ik}$ is the fraction of annotators selecting response $k$ for triplet $i$, and $p_{ik}$ is its predicted probability.

\begin{table}[t]
\centering
\fontsize{9}{11}\selectfont
\setlength{\tabcolsep}{3pt}
\renewcommand{\arraystretch}{1}
\caption{Baselines grouped by encoder and training objective. Fused and hidden cosine use embeddings before and after the downstream hidden layer.
VAD distance is the negative Euclidean distance between predicted VAD vectors.
$^{*}$ indicates a significant difference from SES-Judge (paired triplet bootstrap for Spearman; exact two-sided McNemar for accuracy, $\alpha=0.05$).}
\label{tab:baselines}
\begin{tabular*}{\columnwidth}{@{\extracolsep{\fill}}>{\raggedright\arraybackslash}p{0.50\columnwidth}rr@{}}
\hline
\rule[-3pt]{0pt}{13pt}Method & Spearman & Accuracy (\%) \\
\hline
\rule[-3pt]{0pt}{13pt}\textbf{SES-Judge} & $\mathbf{0.5297}^{\phantom{*}}$ & $\mathbf{74.96}^{\phantom{*}}$ \\
\hline
\multicolumn{3}{@{}l}{\rule[-3pt]{0pt}{13pt}\textit{\textbf{Embedding cosine similarity}}} \\
\rule[-3pt]{0pt}{13pt}\hspace{0.6em}EmotionRankCLAP & $0.3537^{*}$ & $63.46^{*}$ \\
\rule[-3pt]{0pt}{13pt}\hspace{0.6em}WavLM-SER & $0.3591^{*}$ & $64.25^{*}$ \\
\rule[-3pt]{0pt}{13pt}\hspace{0.6em}emotion2vec & $0.3064^{*}$ & $65.51^{*}$ \\
\hline
\multicolumn{3}{@{}l}{\rule[-3pt]{0pt}{13pt}\textit{\textbf{WavLM-SER + downstream training}}} \\
\rule[-3pt]{0pt}{13pt}\hspace{0.6em}VAD reg. / Fused cosine & $0.4820^{*}$ & $71.65^{*}$ \\
\rule[-3pt]{0pt}{13pt}\hspace{0.6em}VAD reg. / Hidden cosine & $0.3660^{*}$ & $64.72^{*}$ \\
\rule[-3pt]{0pt}{13pt}\hspace{0.6em}VAD reg. / VAD distance & $0.3655^{*}$ & $66.30^{*}$ \\
\rule[-3pt]{0pt}{13pt}\hspace{0.6em}Classification / Fused cosine & $0.4594^{*}$ & $70.55^{*}$ \\
\rule[-3pt]{0pt}{13pt}\hspace{0.6em}Classification / Hidden cosine & $0.3360^{*}$ & $61.57^{*}$ \\
\rule[-3pt]{0pt}{13pt}\hspace{0.6em}Classification / Negative KL & $0.3227^{*}$ & $59.53^{*}$ \\
\rule[-3pt]{0pt}{13pt}\hspace{0.6em}VAD triplet / Cosine & $0.4736^{*}$ & $68.35^{*}$ \\
\hline
\multicolumn{3}{@{}l}{\rule[-3pt]{0pt}{13pt}\textit{\textbf{emotion2vec + downstream training}}} \\
\rule[-3pt]{0pt}{13pt}\hspace{0.6em}VAD reg. / Fused cosine & $0.4258^{*}$ & $68.82^{*}$ \\
\rule[-3pt]{0pt}{13pt}\hspace{0.6em}VAD reg. / Hidden cosine & $0.3399^{*}$ & $66.93^{*}$ \\
\rule[-3pt]{0pt}{13pt}\hspace{0.6em}VAD reg. / VAD distance & $0.3174^{*}$ & $64.72^{*}$ \\
\rule[-3pt]{0pt}{13pt}\hspace{0.6em}Classification / Fused cosine & $0.4124^{*}$ & $66.46^{*}$ \\
\rule[-3pt]{0pt}{13pt}\hspace{0.6em}Classification / Hidden cosine & $0.3063^{*}$ & $62.20^{*}$ \\
\rule[-3pt]{0pt}{13pt}\hspace{0.6em}Classification / Negative KL & $0.2903^{*}$ & $62.20^{*}$ \\
\rule[-3pt]{0pt}{13pt}\hspace{0.6em}VAD triplet / Cosine & $0.3671^{*}$ & $67.09^{*}$ \\
\hline
\multicolumn{3}{@{}l}{\rule[-3pt]{0pt}{13pt}\textit{\textbf{LALMs as judges}}} \\
\rule[-3pt]{0pt}{13pt}\hspace{0.6em}Gemini 3.8 Flash & $0.3813^{*}$ & $60.63^{*}$ \\
\rule[-3pt]{0pt}{13pt}\hspace{0.6em}Qwen3-Omni & $0.0739^{*}$ & $3.15^{*}$ \\
\hline
\end{tabular*}
\end{table}

\begin{table}[t]
\centering
\fontsize{9}{11}\selectfont
\setlength{\tabcolsep}{3pt}
\renewcommand{\arraystretch}{1}
\caption{Comparison with classification baselines trained on MSP-Podcast and synthetic speech from SES-Bench Train.
Fused and hidden cosine use embeddings before and after the downstream hidden layer.
$^{*}$ indicates a significant difference from SES-Judge (paired triplet bootstrap for Spearman; exact two-sided McNemar for accuracy, $\alpha=0.05$).}
\label{tab:synthetic_classification}
\begin{tabular*}{\columnwidth}{@{\extracolsep{\fill}}>{\raggedright\arraybackslash}p{0.50\columnwidth}rr@{}}
\hline
\rule[-3pt]{0pt}{13pt}Method & Spearman & Accuracy (\%) \\
\hline
\rule[-3pt]{0pt}{13pt}\textbf{SES-Judge} & $\mathbf{0.5297}^{\phantom{*}}$ & $\mathbf{74.96}^{\phantom{*}}$ \\
\hline
\multicolumn{3}{@{}l}{\rule[-3pt]{0pt}{13pt}\textit{\textbf{WavLM-SER + downstream training}}} \\
\rule[-3pt]{0pt}{13pt}\hspace{0.6em}Classification / Fused cosine & $0.3559^{*}$ & $66.61^{*}$ \\
\rule[-3pt]{0pt}{13pt}\hspace{0.6em}Classification / Hidden cosine & $0.1783^{*}$ & $57.80^{*}$ \\
\rule[-3pt]{0pt}{13pt}\hspace{0.6em}Classification / Negative KL & $0.1330^{*}$ & $56.38^{*}$ \\
\hline
\multicolumn{3}{@{}l}{\rule[-3pt]{0pt}{13pt}\textit{\textbf{emotion2vec + downstream training}}} \\
\rule[-3pt]{0pt}{13pt}\hspace{0.6em}Classification / Fused cosine & $0.3936^{*}$ & $66.14^{*}$ \\
\rule[-3pt]{0pt}{13pt}\hspace{0.6em}Classification / Hidden cosine & $0.2257^{*}$ & $59.53^{*}$ \\
\rule[-3pt]{0pt}{13pt}\hspace{0.6em}Classification / Negative KL & $0.1557^{*}$ & $56.85^{*}$ \\
\hline
\end{tabular*}
\end{table}

\section{Experiment}
\label{exp}

\subsection{Setup}
We train SES-Judge on all SES-Bench Train comparisons, regardless of agreement among annotators (Section~\ref{sec:annotation}).
For evaluation, we map A3, A2, A1, SAME, B1, B2, and B3 to the discrete scores $3,2,1,0,-1,-2,-3$, respectively.
Positive scores favor A, while negative scores favor B.
We compute Spearman correlation between model predictions and mean human ratings over all 848 test comparisons.
These ratings capture both preference direction and strength.

We also report accuracy on a simplified A/B choice task using 635 comparisons, excluding comparisons with majority SAME responses or No Majority.
Score-based methods select the candidate with the higher similarity score, while LALM predictions follow the response aggregation described in Section~\ref{sec:baselines}.
Tied similarity scores and LALM predictions of SAME are counted as incorrect.

\subsection{Baselines}
\label{sec:baselines}
Table~\ref{tab:baseline_summary} summarizes models, training objectives, and scoring methods.
We evaluate cosine similarity using pretrained embeddings from EmotionRankCLAP~\cite{chandra25_interspeech}, WavLM-SER, and emotion2vec~\cite{ma-etal-2024-emotion2vec}.
For emotion classification and methods based on valence, arousal, and dominance (VAD), we use frozen WavLM-SER and emotion2vec base encoders with the same per-layer projection, normalization, and learnable weighted sum as SES-Judge, producing a fused representation.
These models are trained on MSP-Podcast's official training split.
For emotion classification only, we exclude utterances with emotion-category labels O (other) or X (no agreement among annotators).
We train classifiers on MSP-Podcast plus all synthetic candidates from SES-Bench Train, assigning each candidate its reference's emotion category.
With both methods using these synthetic utterances, we compare supervision from emotion categories and human similarity judgments.
Following emotion2vec's downstream head design~\cite{ma-etal-2024-emotion2vec}, we use a 256-unit linear layer with ReLU followed by an eight-class output layer.
For VAD regression, we retain the hidden layer and use three sigmoid outputs for VAD ratings scaled to $[0,1]$.
Triplet learning uses VAD distances to select positive and negative utterances and trains a final linear projection of the fused representation, using normalized embeddings.

For Qwen3-Omni~\cite{xu2025qwen3} and Gemini 3.8 Flash\footnote{\url{https://ai.google.dev/gemini-api/docs/models/gemini-3.8-flash}}, we ask ``Which of A and B is closer to the REFERENCE in emotion?'' with the seven human annotation options.
Each model answers ten times per triplet, with five presentations in each A/B order.
We restore candidate identities and average signed responses for Spearman correlation.
For accuracy, we group responses into A, B, or SAME and take the most frequent category, assigning SAME when the highest vote counts are tied.

\subsection{Experimental Results}
Table~\ref{tab:baselines} compares SES-Judge with methods trained without SES-Bench's comparison labels.
SES-Judge achieves a Spearman correlation of 0.5297, significantly outperforming all baselines, and the highest accuracy of 74.96\%.
The correlation advantage extends across embedding metrics, models trained with emotion labels, and prompted LALMs.
Qwen3-Omni's low accuracy primarily reflects frequent SAME predictions, which account for 94.96\% of its aggregated predictions on the A/B evaluation subset.
Together, these results support using SES-Bench's comparisons to learn a metric aligned with human judgments of preference direction and strength.

Across both encoders, VAD regression yields higher Spearman correlation and accuracy than emotion classification across the three readouts.
For both training objectives, cosine similarity between fused representations performs best, outperforming hidden-layer cosine similarity and output-space distances.
These results suggest that representations trained with emotion labels retain similarity information that is less accessible through their downstream heads.

Table~\ref{tab:synthetic_classification} shows that SES-Judge retains its advantage when classifiers also receive the synthetic utterances used in its training.
SES-Judge significantly outperforms the classifiers using either encoder across three readouts on both metrics.
Thus, even with the same synthetic utterances, classifiers trained with emotion category labels do not match SES-Judge's agreement with human judgments.
This finding highlights the value of SES-Bench as a source of human similarity supervision, beyond providing synthetic training speech.

\section{Conclusion}
We introduced SES-Bench, a benchmark for evaluating perceived emotion similarity through graded human comparisons.
Its annotations capture both preference direction and relative strength and can be used to train and evaluate emotion similarity models.
Trained on these comparisons, SES-Judge shows stronger agreement with human judgments than embedding cosine similarity and prompted LALMs.
Future work will explore using SES-Judge as a reward model to guide expressive speech generation toward closer emotional similarity to a reference utterance.

\bibliographystyle{IEEEbib}
\bibliography{strings,refs}

\end{document}